\documentclass{elsart}
\usepackage{graphicx}
\journal{Astroparticle Physics}

\begin{document}

\begin{frontmatter}
\title{The Prompt Lepton Cookbook}
\author{C.G.S. Costa\thanksref{ead}}
\thanks[ead]{Email address: cesar.costa@ulb.ac.be}
\address{Service de Physique Th\'eorique, CP 225, 
Universit\'e  Libre de Bruxelles, 
Boulevard du Triomphe, 1050 Brussels, Belgium}

\begin{abstract}
We review the calculation of the prompt lepton flux, produced in
the atmosphere by the semileptonic decay of charmed particles. 
We describe side by side the intermediary ingredients used by
different authors, which include not only the charm production
model, but also other atmospheric particle showering parameters. 
After evaluating separately the relevance of each single
ingredient, we analyze the effect of different combinations over
the final result.  We highlight the impact of the prompt lepton
flux calculation upon high-energy neutrino telescopes. 
\end{abstract}

\end{frontmatter}

\newpage
\section{Introduction}
Very-high-energy (above 1 TeV)  
neutrino astronomy is currently a subject of  great interest, 
promising to expand our observational range of the
Universe  in an unique way\cite{reports}. Such energetic neutrinos may carry
information from the sources of the highest energy phenomena ever observed in
cosmic rays,  possibly coming from active galactic nuclei (AGN) or
gamma ray bursts (GRB). 
They may probe the early stages of the Universe and its farthest distances. 
In addition, they will contribute to the search of weakly interacting massive 
particles (WIMP), supernova explosions, monopoles, besides the discovery
potential for new physics, which we can even not imagine yet. 
Neutrino telescopes under development, like the Antarctic Muon and Neutrino
Detector Array (AMANDA) and the experiment at Lake Baikal, 
are already operational, and producing their first 
results\cite{amanda:ap00,baikal:ap00}. In addition, great activity 
is planned for the near future\cite{spiering:nu00}.   

Aside from these perspectives, the operation of a neutrino detector 
at energies above 1 TeV poses challenging difficulties. One of the major
limitations in the detection of a cosmic high-energy neutrino (from galactic 
or extra-galactic origin) is the background from atmospheric muons and
neutrinos (produced by the interaction of high-energy cosmic rays in the
atmosphere). 

The source of the atmospheric neutrino background changes with energy, 
in a way governed by the critical energy $\varepsilon_{critic}$  
of the parent particle. This is the energy for which 
the decay and interaction lengths are equal. 
Above this energy the parent particle is likely to interact 
or be slowed down before decaying into a neutrino.  
The critical energy is calculated in terms of the particle rest 
energy $mc^2$, the mean life $\tau$ and, by adopting 
the isothermal atmosphere approximation, a scale constant 
$h_{o}$\cite{gaisser:92}:
\begin{equation}
\varepsilon_{critic} \: = \: \frac{mc^2}{c\tau} \: h_{o}. 
\label{eq:epsilon} 
\end{equation}

\noindent Table~\ref{table:pdg} summarizes several particle properties, 
derived from the Review of Particle Physics~\cite{pdg:00}. 
Comparing the critical energies we observe that muon decays 
contribute substantially to the atmospheric  
lepton flux only up to a few GeV's, while the decays 
in flight of charged pions and kaons are still 
important up to about 1-10 TeV. They give rise to the  ``conventional''
atmospheric lepton flux. Above this energy range, the semileptonic decay of
very-short lived charmed particles (mainly $D$-mesons and 
$\Lambda_{c}^{+}$-hyperons) 
is the dominant source, despite their low production rate. 
The main contribution comes from the decay modes 
\[  D \rightarrow K + \mu + \nu \:\: {\rm and} \:\: 
  \Lambda_{c} \rightarrow \Lambda_{o} + \mu + \nu . \] 

\noindent Muons and neutrinos thus generated are 
called ``prompt leptons'', and they exhibit a flatter 
(and thus harder) energy spectrum. 
The lack of precise information on high-energy charm production
in hadron-nucleus collisions leads to a great uncertainty in the estimate 
of the leptonic flux above 100~TeV. In addition, different authors 
do not use the same atmospheric particle showering routines,
turning the comparison even more difficult. 
The predictions of resulting fluxes span up to four orders of
magnitude!  

It is our purpose to bring some light to the forum of prompt
lepton fluxes, describing side by side the many ingredients of the
calculation. After evaluating separately the relevance of each
single shower parameter, we analyze the effect of different
combinations over the final result. 
We investigate the fluxes
of $\mu$, $\nu_{\mu}$ and $\nu_{e}$, leaving the case of $\tau$ and 
$\nu_{\tau}$ for a further work.

\begin{table}[hb]  
\caption[]{\label{table:pdg} Particle data summary.}    
\smallskip
\centering
\def\arraystretch{1.25}
\begin{tabular}{lcrccr}     
\hline\hline  
Particle &Elementary &$mc^2$ 
&$c\tau$ &$\varepsilon_{critic}$$^{(1)}$ &Branching \\ 
&contents &(MeV) &&(GeV) &ratio $B_{i}$$^{(2)}$ \\ \hline
$D^{+},D^{-}$ &$c\bar{d},\bar{c}d$ &1870 &317 $\mu$m 
&$3.8 \times 10^7$ &17.2 \%\\ 
$D^{o},\bar{D}^{o}$ &$c\bar{u},\bar{c}u$   &1865 &124 $\mu$m 
&$9.6 \times 10^7$ &6.8 \%\\ 
$D_{s}^{+},D_{s}^{-}$ &$c\bar{s},\bar{c}s$ &1969 &149 $\mu$m
&$8.5 \times 10^7$ &5.2 \%\\  
$\Lambda_{c}^{+}$  &$udc$  &2285 &62 $\mu$m &$2.4 \times 10^8$ & 4.5 \%\\ 
\hline    
$\mu^{+},\mu^{-}$ &lepton              &106 &659 m &1.0 &100 \%  \\  
$\pi^{+},\pi^{-}$ &$u\bar{d},\bar{u}d$ &140 &7.8 m &115 &100 \%  \\
$K^{+},K^{-}$ &$u\bar{s},\bar{u}s$     &494 &3.7 m &855 &63.5 \% \\
$\Lambda^{o}$  &$uds$  &1116 &7.9 cm &$9.0 \times 10^4$ &0.1 \% \\
\hline\hline 
\multicolumn{6}{l}{\footnotesize  
 (1) According to Eq.~(\ref{eq:epsilon}), with $h_{o}=6.4$ km.} \\
\multicolumn{6}{l}{\footnotesize 
 (2) For inclusive decays yielding leptons.}
\end{tabular}  
\end{table}

\section{Calculation of the prompt lepton flux}
\label{section:calculation} 
The calculation of the prompt lepton flux has been carried 
out in the past (see, e.g. Ref.\cite{bugaev:98} and references
therein), mainly with the purpose to investigate the effects of
choosing a different charm production model.  
The outline of the calculation is basically the same in all of 
these works.  We start from the primary cosmic ray flux at the 
top of the atmosphere, with a composition supposed 
to be dominated by protons, and  evaluate the flux of nucleons at any 
atmospheric depth. Those nucleons interact with the nuclei of air 
to produce secondary particle showers. 
For energies above a 
few TeV, the  secondary particles of interest to be followed are 
the charmed hadrons, for they will be the main source of 
high energy atmospheric leptons. 
We will, therefore, consider the contribution of mesons $D^{\pm}, D^{o}$ 
and $\bar{D}^{o}$ (comprising in the same notation the overall
contribution  of $D$ and $D^{*}$ mesons), of the strange
$D_{s}^{\pm}$ and of the $\Lambda_{c}^{+}$-hyperon. 

Because the production rate of the mesons $D_{s}$ is relatively
lower (about $20 \%$ of the $D$ production cross section), some 
authors neglect their contribution, although their critical
energy and branching ratio are comparable to those of the other
charmed particles (see Table~\ref{table:pdg}).
When calculating the flux of $\tau$ and $\nu_{\tau}$, which we do not
consider in this analysis, the role of $D_{s}$ turns out to be
mostly important\cite{pr:99}.

It is straightforward to calculate the flux of charm particles 
at any depth, and they will promptly decay yielding electrons, muons 
and neutrinos. 
We may integrate the flux for all possible charmed parent 
production and decay depths, and for all possible production and decay 
energies, leading to the flux of a chosen lepton at a given depth and 
energy. 

For detailed calculations we refer to 
Refs.\cite{bugaev:98,volkova:87,halzen:93}, and we follow 
the notation of the latter, to present here only main results. 
Let's write the primary cosmic-ray spectrum as a power law in energy:
\begin{equation}
\Phi_{N} (E_{N},x=0) \: = \: N_{o} \:
E_{N}^{-(\gamma+1)}, 
\label{eq:Fprime}
\end{equation}

\noindent where $\Phi_{N}(E_{N},x)$, given in
(GeV.cm$^2$.s.sr)$^{-1}$, 
is the differential flux of nucleons with energy $E_{N}$, in GeV, 
and $x$ is the slant depth penetrated by the cascade, measured in 
g/cm$^2$ from the top of the atmosphere ($x=0$) downward along 
the direction of the incident nucleon. The constant $N_{o}$ is 
the amplitude, or differential spectrum normalization; and 
$\gamma$ is the spectral index, or slope of the integral primary 
spectrum. 

After developing to a certain depth $x$, the nucleonic flux is given 
in terms of $\Lambda_{N}$, the nucleonic attenuation length\cite{gaisser:92}: 
\begin{equation}
\Phi_{N}(E_{N},x,\theta)\: = \: N_{o} \:
E_{N}^{-(\gamma+1)} \:
e^{-x/\Lambda_{N}}.  
\label{eq:Fnucx}
\end{equation}

The resulting flux of secondary particles of type-$i$ 
($i = D^{\pm}, D^{o}, \bar{D}^{o},  D_{s}^{\pm}, \Lambda_{c}^{+}$) 
is calculated by convolution of the nucleonic flux with the
production spectrum of secondary particles:  
\begin{equation}
\Phi_{i}(E_{i},x,\theta)
\: = \: K_{i} (E_{i},\gamma) \: \int_{0}^{x} dx' \: 
\left(\frac{x'}{x}\right)^\eta \: 
\exp\left\{-\frac{(x-x')}{\lambda_{i}} \:-\frac{x'}{\Lambda_{N}}\right\}, 
\label{eq:Fix}
\end{equation}

\noindent defining 
\[ \eta \: = \: \frac{\varepsilon_{critic}}{E_{i}\:\cos\!\theta}, \]

\noindent where the dependence in the zenith angle holds for 
$\theta \leq 60^{o}$. For higher zenith angles the curvature of 
Earth must be taken into account. Both the nucleonic attenuation 
length $\Lambda_{N}$ and the charmed particle interaction length $\lambda_{i}$ 
are given in units g/cm$^2$. 
The production spectrum of charmed particles, 
weighted by the primary nucleonic spectrum, is written as:
\begin{eqnarray}
K_{i} (E_{i},\gamma) & = & 
\int_{E_{i}}^{\infty} dE_{N} \: \: \frac{N_{o}}{\lambda_{N}} \:
E_{N}^{-(\gamma+1)} \:  
\frac{dW^{Ni}(E_{i},E_{N})}{dE_{i}}, 
\label{eq:Ki}\\ 
& = & \frac{N_{o}}{\lambda_{N}} \:
E_{i}^{-(\gamma+1)} \: Z_{Ni}(\gamma). 
\label{eq:KiZi}
\end{eqnarray}

In this notation, $dW^{Ni}/dE_{i}$ denotes the energy distribution of 
secondary particles, and represents the probability that a particle 
of type-$i$ with energy $E_{i}$ is created in the interaction of 
an incident nucleon $N$ of energy $E_{N}$ with an air nucleus. 
It is directly related to the inclusive cross section for
secondary particle production. 
Eq.~(\ref{eq:KiZi}) is obtained assuming a mild energy
dependence for the nucleonic interaction length $\lambda_{N}$, and
defining $Z_{Ni}(\gamma)$, the particle production spectrum-weighted 
moment\cite{gaisser:92}, also called production ``Z-moment'': 
\begin{equation}
Z_{Ni}(\gamma)  =  
\int_{0}^{1} x_{F}^{\! \gamma} 
\left( \frac{dW^{Ni}}{dx_{F}}\right) \: dx_{F} ,
\label{eq:Zch}
\end{equation}

\noindent where $x_{F}$ is the Feynman variable $x_{F} \equiv
p_{L}/p_{L}^{max}$, with $p_{L}$ as the produced particle
longitudinal momentum. At the high-energy limit the
Feynman-$x$ also represents the ratio of  the final particle
energy to the incident particle energy, $x_{F}=E_{i}/E_{N}$ 
(beware confusion with atmospheric depth $x$). 

In order to evaluate the flux $\Phi_{l}(E_{l},x,\theta)$ of
leptons  ($l=\mu$ or $\nu$), with energy $E_{l}$ and  zenith angle
$\theta$ at depth $x$, we need to fold the energy  distribution
$df^{l}/dE_{l}$ of the produced lepton with the spectrum  
$D_{i}(E_{i},x'',\theta)$
of decaying parents, for any decay depth $x''$   
and any available parent energy $E_{i}$: 
\begin{equation}
\Phi_{l}(E_{l},x,\theta)
         \:=\: \int_{0}^{x} dx'' 
         \:\: \int_{E_{i}^{max}}^{E_{i}^{min}} dE_{i} 
         \:\: \frac{df^{l}}{dE_{l}} \:\: 
         D_{i}(E_{i},x'',\theta),
\label{eq:Fnil}
\end{equation}

\noindent with  
\begin{eqnarray}
D_{i}(E_{i},x'',\theta)
&=& B_{i} \: \frac{1}{d_{i}} \: 
\Phi_{i}(E_{i},x'',\theta) , 
\label{eq:Di}\\
d_{i} &=& \frac{x''}{\eta} 
\:=\: \frac{x'' \cos\!\theta \:E_{i}}{\varepsilon_{critic}}
\label{eq:decay}
\end{eqnarray}

\noindent where $B_{i}$ is the branching ratio yielding 
leptons in the parent-$i$ decay (see Table~\ref{table:pdg}), 
and $d_{i}$ is the particle-$i$ decay length. 

The muon and neutrino production energy distributions 
$df^{l}/dE_{l}$ used in  Eq.~(\ref{eq:Fnil}) are given by the 
semileptonic three-body decay phase space integrals, obtained
from kinematics con\-sid\-er\-ations\cite{gaisser:92,hagedorn:63}. 
Some authors\cite{lipari:93,tig:96,prs:99} define a decay
Z-moment in analogy to the production Z-moment, Eq.~\ref{eq:Zch}, 
just replacing $dW^{Ni}/dE_{i}$ by  $df^{l}/dE_{l}$. 
Making use of this definition, it is possible to write an
approximate solution (valid for energies $E_{l}
<\varepsilon_{critic}$) exploring the fact that the critical
energy for charmed particles is very high. Nevertheless, in the present
work we will use the complete solution for calculating the prompt
lepton flux, given by the set of Equations~(\ref{eq:Fix}) to
(\ref{eq:decay}).

\section{Showering Process}
\label{section:showering}
Once the calculation is established, the next step is to choose
the ingredients that characterize the showering process in the
atmosphere. We will browse through the literature to extract
different parametrizations to be compared. The main parameters to
define are: the primary spectrum normalization and slope, the
nucleonic and charm interaction lengths, the nucleonic attenuation
length and the charm production Z-moment. 

\subsection{Primary spectrum}
\label{section:primes}
The primary cosmic ray flux at the top of the atmosphere,
Eq.~(\ref{eq:Fprime}), can be rewritten as to incorporate the
effect of the change in slope (``knee'') observed in the energy spectrum, 
at energy $E_{knee}$:
\begin{eqnarray}
\Phi_{N} (E_{N}, x=0) & = & N_{1} \: E_{N}^{-(\gamma_{1}+1)}, 
\:\: E_{N} <E_{knee} \; \mbox{{\rm (in GeV)}}; 
\nonumber \\ 
 & = & N_{2} \: E_{N}^{-(\gamma_{2}+1)}, \:\: E_{N} >E_{knee} 
\; \mbox{{\rm (in GeV)}}.
\label{eq:primes}
\end{eqnarray}

\begin{figure} 
\begin{center}
 \includegraphics[height=11cm,angle=-90]{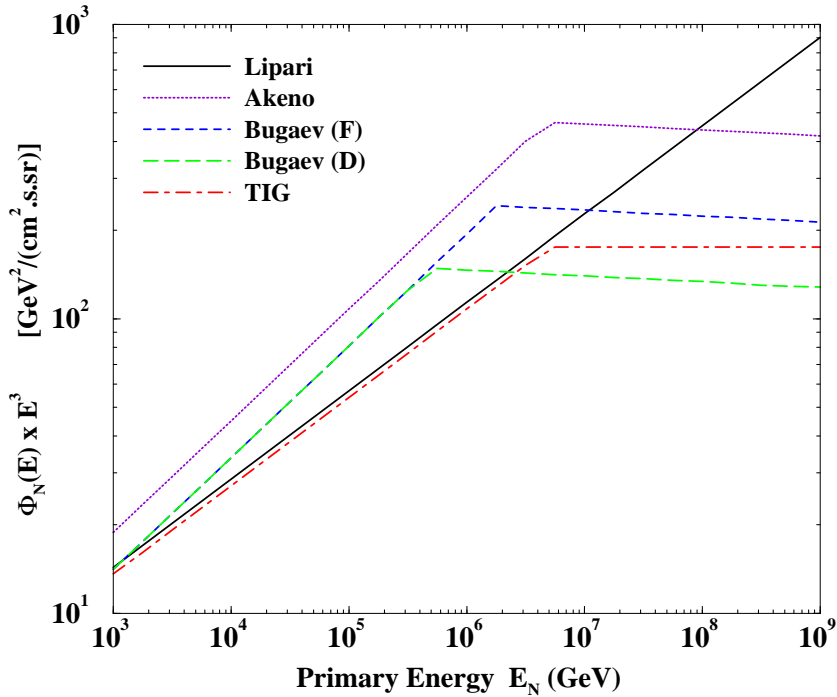}
\end{center}
\caption[]{Comparison of primary cosmic-ray energy spectra, as  
given by different parametrizations (see Table~\ref{table:primes}).
The primary flux is multiplied by $E_{N}^{3}$, 
so that the structure at the knee can be better appreciated, 
and the different parametrizations compared.}
\label{fig:primes}   
\end{figure}

Table~\ref{table:primes} indicates some typical values for the
parameters  in Eq.~(\ref{eq:primes}), used by different authors.
Lipari\cite{lipari:93} quotes a parametrization consistent with
both the JACEE  balloon borne ex\-per\-i\-ments\cite{jacee:90} and
the values given by Gaisser\cite{gaisser:92}, adopting a single
slope, because  his analysis is mainly aimed at energies below the
knee. The AKENO experiment\cite{akeno:84} obtained a description
of the primary spectrum, covering the knee region, from data on
size spectra of electrons and muons at high energy.  
Bugaev {\it et al.}\cite{bugaev:89} use a semiempirical model
which takes into account detailed chemical composition of the
primary spectrum, translated here in terms of 
Eq.~(\ref{eq:primes}). They propose two options (Model F and
Model D), differing on the hypothesis responsible for the change
in slope at the knee. Thunman, Ingelman and Gondolo (TIG)
\cite{tig:96} follow the JACEE trend   below the knee. 
Above the knee they adopt the same slope as  Volkova {\it et
al.}\cite{volkova:87},  for which no normalization constant is
reported, due to their interest limited to flux ratios.
Figure~\ref{fig:primes} displays the energy spectra corresponding
to those parametrizations.

\begin{table}  
\caption[]{\label{table:primes} 
Primary cosmic-ray flux$^{(1)}$} 
\smallskip
\centering
\def\arraystretch{1.25}
\begin{tabular}{lrrrrr} 
\hline\hline 
Label  &\multicolumn{2}{c}{$E<E_{knee}$}    
     &$E_{knee}$ &\multicolumn{2}{c}{$E>E_{knee}$} \\  
       &$N_{1}$ &$\gamma_{1}$ &(GeV) 
       &$N_{2}$ &$\gamma_{2}$ \\
\hline 
Lipari &1.70 &1.70 & - & - & - \\
Akeno       &1.35 &1.62 &$4.67 \times 10^{6}$ &630 &2.02\\
Bugaev (F) &1.02  &1.62 &$1.9  \times 10^{6}$ &323 &2.02\\
Bugaev (D) &1.02  &1.62 &$5.2  \times 10^{5}$ &193 &2.02\\
TIG &1.70 &1.70 &$5 \times 10^{6}$ &174 &2.00\\
\hline\hline   
\multicolumn{6}{l}{\footnotesize 
(1) According to Eq.~(\ref{eq:primes}),}\\
\multicolumn{6}{l}{\footnotesize 
 with $N_{1}$ and $N_{2}$ given 
in units (GeV.cm$^2$.s.sr)$^{-1}$.}
\end{tabular}  
\end{table}  

\subsection{Interaction and attenuation lengths}
\label{section:sigmas}
The nucleonic interaction length, $\lambda_{N}$, represents the
mean free path of nucleons in the atmosphere (given in g/cm$^2$). 
It is related to $\sigma_{in}^{\! N-air}$, the total inelastic
cross section for  collisions of nucleons $(N)$ with air nuclei,
through the relation    
\begin{equation}
\lambda_{N} (E)  =  \frac{A. m_{p}}{\sigma_{in}^{\! N-air} (E)} 
           =  \frac{24100}{\sigma_{in}^{\! N-air} (E)}, 
\label{eq:lnuc}
\end{equation}

\noindent where we used the average atomic number for air nuclei $A$=14.5,
$m_{p}$ is the proton mass and $\sigma_{in}^{\! N-air}$ must be given in $mb$.
There are different parametrizations to the inelastic $N-$air
cross section as a function of energy. Some authors considered it to be 
constant\cite{volkova:87,lipari:93}; others to be rising with
energy: as a power-law\cite{akeno:84}, as a logarithmic
dependence\cite{bugaev:98,bugaev:89}, or as a log-squared
dependence\cite{chs:95,bhs:00}.

The attenuation length, $\Lambda_{N}$, which governs the exponential decay 
of the nucleonic flux with increasing depth, see Eq.~(\ref{eq:Fnucx}), 
represents the net effect in the interplay between interaction losses 
and regeneration of the number of nucleons in the cascade development, 
and is given by  

\begin{equation}
\Lambda_{N} \: = \: \frac{\lambda_{N}}{\left( 1-Z_{NN}(\gamma) \right)}. 
\label{eq:Attnuc}
\end{equation}

\noindent The interaction length $\lambda_{N}$ dictates the
losses and the nucleon-to-nucleon  spec\-trum-weighted Z-moment
$Z_{NN}(\gamma)$ accounts for the survival  rate of nucleons.
$Z_{NN}$ is calculated in  analogy to Eq.~(\ref{eq:Zch}), with the
outgoing particles being the  regenerated nucleons. Some authors
use an approximately constant
value\cite{bugaev:98,lipari:93,bhs:00}, others adopt
Feynman-scaling with a change in value at the knee
energy\cite{gaisser:92,chs:95}, while in Ref.~\cite{tig:96} it is
assumed the   violation of Feynman-scaling. 

The charm particle of type-$i$ may have an interaction length 
$\lambda_{i}$ in the atmosphere calculated analogously as for the nucleons, 
Eq.~(\ref{eq:lnuc}), substituting $\sigma_{in}^{\! N-air}$ by 
$\sigma_{in}^{\! i-air}$.
As in the parametrization of the nucleonic inelastic cross
section, we find authors adopting a charm cross section which is
constant\cite{volkova:85}, or which increases with energy either
as a power-law\cite{mm:86}, or logarithmically\cite{bugaev:98}.

\subsection{Charm production spectrum-weighted moments}
\label{section:zch}
The key information for the evaluation of the prompt lepton flux is the 
behavior of the charm spectrum-weighted moments, given by a specific 
charm production model. 
Among these models, we select three different ones, to be compared
in the present study. 

\begin{figure} 
\begin{center}
 \includegraphics[height=11cm,angle=-90]{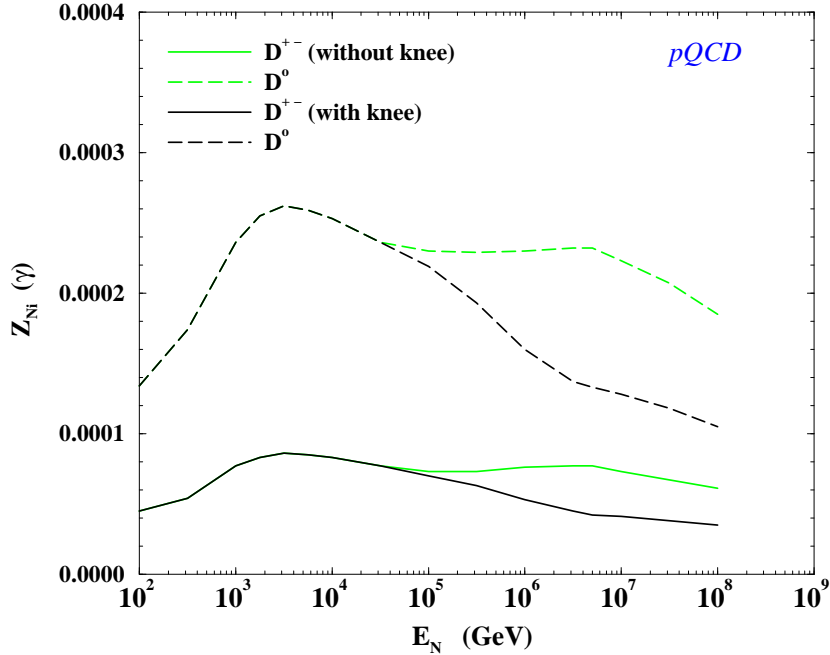}
\end{center}
\caption[]{Charm Z-moment components for pQCD, 
considering the  primary spectrum either with or without a knee, 
as a function of  energy.}
\label{fig:pqcd}   
\end{figure}

\begin{description}
\item[QGSM:] 
Quark Gluon String Model, a semiempirical model of charm
production based on the non-perturbative QCD calculation by
Kaidalov and Piskunova\cite{kaidalov:86}, normalized to
accelerator data, and applied to the prompt muon calculation by
Volkova {\it et al.}\cite{volkova:87}.

The Z-moments are calculated numerically from  Eq.~(\ref{eq:Zch}), with

\begin{equation}
\frac{dW^{Ni}}{dx_{F}} =   
 \frac{\sigma_{NA}^{\! i}}{\sigma_{in}^{\! N-air}} 
                 \: \frac{df_{i}}{dx_{F}}, 
\label{eq:Zch1}
\end{equation}

\noindent where the particle production spectrum is parametrized
by: 
\begin{eqnarray}
\frac{df_{i}}{dx_{F}} &=& \:\:\:\
\frac{1.08}{x_{F}} \:(1-x_{F})^{5}
   \:\:\:\mbox{{\rm for $D$ production}},
\nonumber \\  &=& \:\:\:\
1.4 \:(1-x_{F})^{0.4}  
   \:\:\: \mbox{{\rm for $\Lambda_{c}$ production}}. 
\end{eqnarray}

\noindent For the total inelastic cross section,
$\sigma_{in}^{\! N-air}$, we used a  parametrization with a
$\log^{2}$ energy dependence\cite{chs:95}. 
Following Ref.\cite{volkova:87}, a mild $\log$ energy dependence
is assumed for the inclusive cross section of charm production,
$\sigma_{NA}^{\! i }$, and the production of $D_{s}$ is neglected. 
Figure~\ref{fig:qgsm} displays the curves of Z-moments for the
different charm components. 

\medskip
\item[RQPM:]  Recombination Quark Parton Model, a
phenomenological non-per\-turb\-a\-tive approach, 
taking into account the contribution of the intrinsic charm to
the production process, in which a $c\bar{c}$ pair is coupled to
more than one constituent of the projectile hadron, as described
by Bugaev {\it et al.}\cite{bugaev:98,bugaev:89}. 

Supposing Feynman-scaling to hold ({\bf RQPM-FS}), the charm
production Z-moments are given simply by:
\begin{equation}
Z_{Ni}(\gamma) \:=\: Z_{\gamma} \:=\: \mbox{\rm constant}, 
\label{eq:Zch2}
\end{equation}

\noindent with parameters for different particles shown in
Table~\ref{table:Zrqpm}.  
Assuming the scaling violation ({\bf RQPM-SV}), the
parametrization turns out to be:  
\begin{equation}
Z_{Ni}(\gamma) 
\:=\: Z_{\gamma} \: \left( \frac{E_{N}}{E_{\gamma}} \right)^{\xi}, 
\label{eq:Zch3}
\end{equation}

\noindent where $\xi= \: 0.177 -0.05 \,\gamma$. 
The parameters are also given in Table~\ref{table:Zrqpm}.  
We obtained the all-charged $D$-meson Z-moment ($i=D^{\pm}$) by averaging 
the individual conjugate moments, the same being done for neutral 
$D$-mesons, ($i=D^{o},\bar{D}^{o}$). Ref.\cite{bugaev:98}
does not take into account the contribution of
$D_{s}$ mesons.  A comparison of the resulting
Z-moments,  with and without scaling, is provided in
Figure~\ref{fig:rqpm}. 

\begin{figure} 
\begin{center}
 \includegraphics[height=11cm,angle=-90]{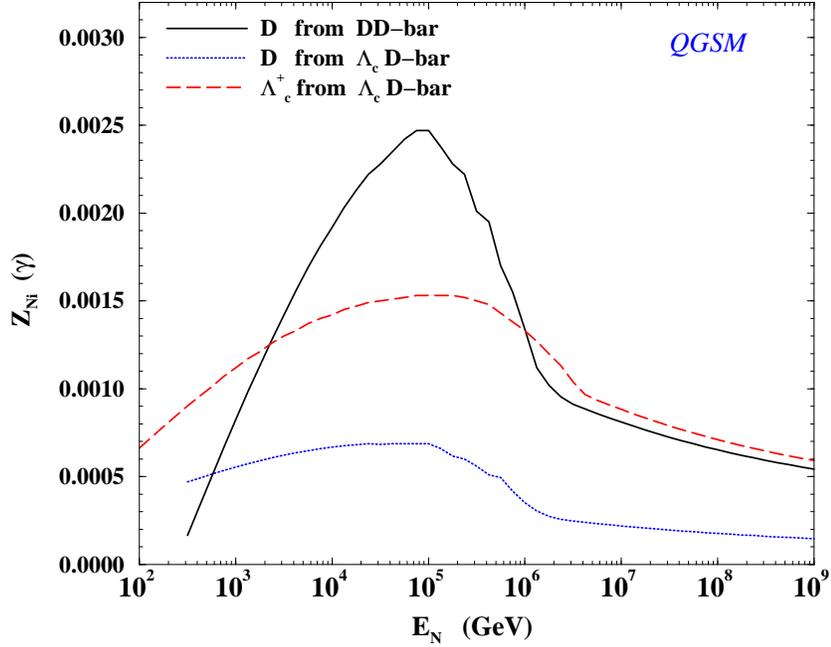}
\end{center}
\caption[]{Charm Z-moment components for QGSM, 
as a function of energy.}
\label{fig:qgsm}   
\end{figure}  

\begin{figure} 
\begin{center}
 \includegraphics[height=11cm,angle=-90]{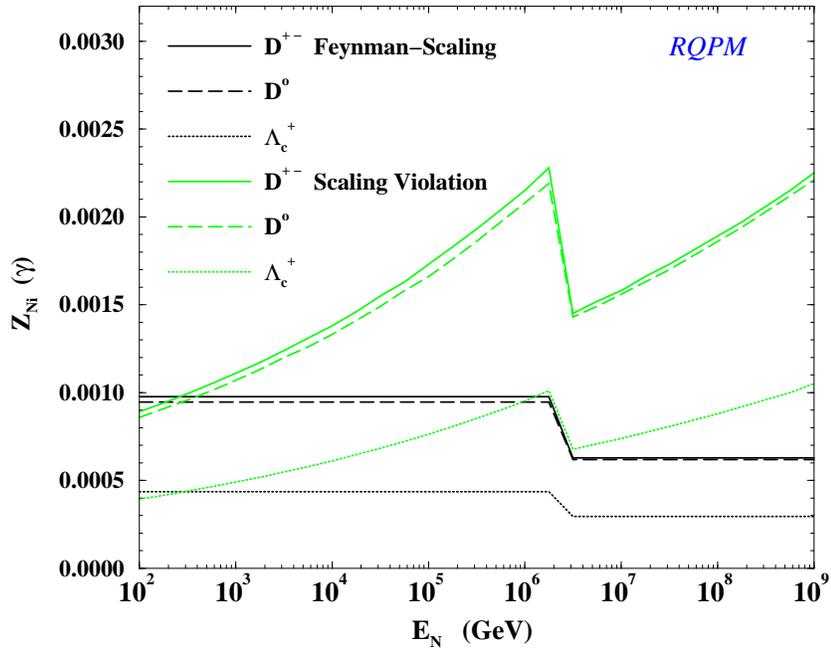}
\end{center}
\caption[]{Charm Z-moment components for RQPM, 
considering respectively Feynman-scaling (FS) and scaling
violation (SV),  as a function of the  energy.}
\label{fig:rqpm}   
\end{figure}  

\begin{table}[ht]  
\caption[]{\label{table:Zrqpm}
RQPM Z-moment parameters $Z_{\gamma}$ $^{(1)}$.}  
\smallskip
\centering
\def\arraystretch{1.25}
\begin{tabular}{lcccrrr} 
\hline\hline 
Label  & $\gamma$ &  $\xi$ & $E_{\gamma}$ (GeV)
& $Z_{\gamma}(D^{\pm})$ & $Z_{\gamma}(D^{o},\bar{D}^{o})$ 
& $Z_{\gamma}(\Lambda_{c}^{+})$
\\ \hline 
RQPM-FS &1.62 & - & - &4.88 $\times 10^{-4}$ &4.73 $\times 10^{-4}$ 
&4.36 $\times 10^{-4}$ \\ 
        &2.02 & - & - &3.14 $\times 10^{-4}$ &3.09 $\times 10^{-4}$ 
&2.95 $\times 10^{-4}$ \\ 
\hline  
RQPM-SV &1.62 &0.096 &$10^{3}$ &5.55 $\times 10^{-4}$ 
&5.35 $\times 10^{-4}$ &4.9 $\times 10^{-4}$ \\ 
     &2.02 &0.076 &$10^{6}$ &6.65 $\times 10^{-4}$ 
&6.55 $\times 10^{-4}$ &6.2 $\times 10^{-4}$ \\ 
\hline\hline   
\multicolumn{7}{l}{\footnotesize 
(1) According to Eqs.~(\ref{eq:Zch2}) and (\ref{eq:Zch3}).} 
\end{tabular}  
\end{table}  

\medskip
\item[pQCD:] 
Perturbative QCD, as calculated by  TIG\cite{tig:96}, using the
Monte Carlo program PYTHIA\cite{pythia:94}, explicitly evaluating
the charm production, up to leading order (LO) in the coupling
constant and including the next-to-leading order (NLO)
distribution effects as an overall factor.  

Figure~\ref{fig:pqcd} presents the curves for pQCD calculations
of the charm production Z-moments carried out supposing a
primary spectrum either with, or without, the knee.
The values of Z-moments for $D$-mesons were extracted directly from 
Ref.\cite{tig:96}. The Z-moments for $\Lambda_{c}^{+}$ 
and $D_{s}$ are derived by taking 
$Z (\Lambda_{c} ) \simeq \: 0.3 \: Z (D)$ and 
$Z (D_{s}) \simeq \: 0.2 \: Z (D)$, respectively, based on values 
assumed for the corresponding cross section ratios. 

More recently Gelmini, Gondolo and Varieschi
(GGV)\cite{ggv:00} updated the calculation to include the full 
contribution of NLO predictions to the lepton fluxes. 
While TIG scales the LO cross sections by a constant factor of $K=2$ 
to obtain the NLO contribution, GGV evaluates explicitly the NLO
component, as tailored by Mangano, Nason and
Ridolfi\cite{mnr:93}. 
At the end, the net calculation corresponds as to multiply the LO term by an
energy dependent factor $K$. In the $10^{2}$ to $10^{11}$ GeV energy interval,
it starts at the lowest energies with $K=3$, decreases to around 2 for most of
the intermediate energies, increasing slightly at the high energy extreme. 
However, the main difference between the two calculations comes from the
extrapolation of the gluon parton distribution function, which produces higher
charm cross sections even at LO. We do not duplicate here the particular
effects implied over the Z-moments, leaving to consider the overall changes,
resulting  from GGV approach, at the evaluation of the lepton fluxes.  
Despite these differences, the atmospheric cascading routines
are the same in both TIG and GGV calculations. 
A discussion of an alternative evaluation of the charm
production based on pQCD, compared to TIG's calculation, is
presented by Pasquali {\it et al.}\cite{prs:99}. 
\end{description} 

In Section~\ref{section:results} we compare the resulting
prompt lepton fluxes, evaluated using each of these models. 
It is interesting to include in this comparison the results
obtained by E. Zas {\it et al.}\cite{halzen:93}. They calculate
extreme cases of charm production, at both low and high production
rate limits.  As for the high end, they assume a charm production
cross section which is 10\% of the total inelastic cross section
(called Model-A), behaving as $\log^2 (s)$ at high energies,
$\sqrt{s}$ being the center-of-mass system (c.m.s.) energy.  At
the low end lies a pQCD model at NLO, with structure functions
given by Kwiecinski-Martin-Roberts-Stirling\cite{kmrs:93},
adopting one choice of parameters that leads to relatively hard
parton distribution functions (called Model-E).

\section{Results}
\label{section:results}
\subsection{High-energy atmospheric muon flux}
\begin{figure} 
\begin{center}
 \includegraphics[height=11cm,angle=-90]{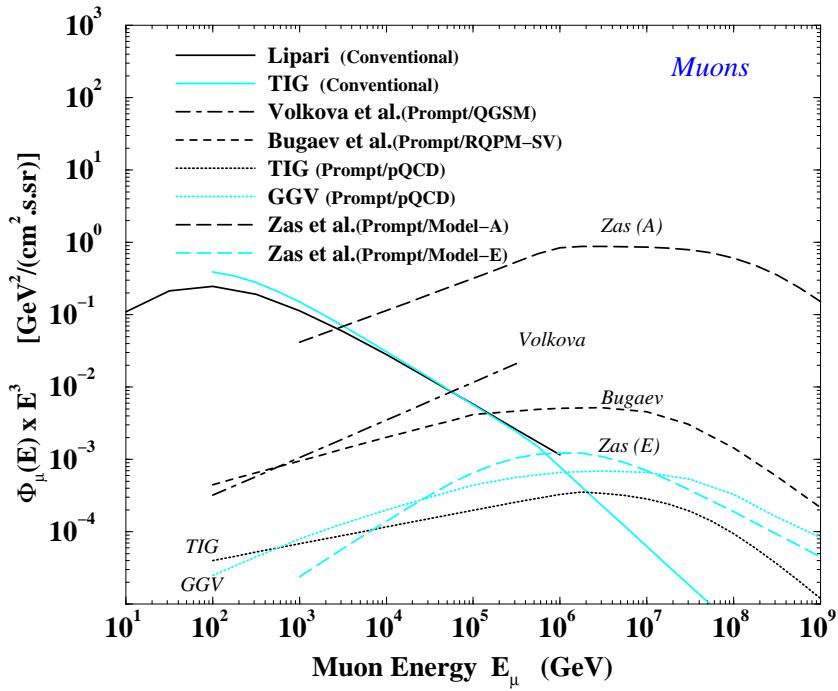}
\end{center}
\caption[]{Comparison of calculated differential vertical muon
fluxes at sea level,  as reported by several authors. The fluxes
are multiplied by $E_{\mu}^{3}$.}
\label{fig:reported}   
\end{figure}

A reproduction of calculated fluxes of high-energy muons
reported by different authors is presented in
Figure~\ref{fig:reported}. Comparing the conventional flux, from
the decay of pions and kaons, obtained by Lipari and TIG, with the
prompt flux, calculated by  Volkova {\it et al.},
Bugaev {\it et al.}, TIG, GGV and Zas {\it et al.}, we note that
the cross-over of the conventional and prompt components may occur 
at energies between $2 \times 10^3$ GeV and $2 \times 10^6$ GeV.
The prompt flux intensity, itself, spans up to four orders of
magnitude! 

We remark that the lowest curves, labeled TIG, GGV and Zas (E),
are calculated assuming the pQCD charm production model. Although
based on the same framework, these calculations of the charm
cross section are subjected to theoretical uncertainties, which
arises from the possible range of charm quark masses, as well as
the factorization and renormalization scale dependence.
Moreover, different assumptions are made for the parton
distribution functions, needed at very small parton momentum
fractions, not measured at accelerators. 
The curve labeled Bugaev is calculated within the RQPM. As an
intrinsic-charm model, it predicts relatively hard inclusive
spectra. In spite of that, the total inclusive cross section can be
made rather large, since it depends strongly on the assumptions
about the charm structure function of the projectile hadron. 
The curve labeled Volkova is obtained from a parametrization given
in Ref.\cite{bugaev:98}, because the original
work\cite{volkova:87} quotes only flux ratios. The QGSM assumed in
this calculation is considered to  represent quite well open-charm
production, and it is known to describe a wide variety of data on
hadronic interactions. However, the model predicts a preferential
production of certain secondary particle species which is not
supported by experiment. In addition, predictions based on this
model seem to exceed the  experimental observations of horizontal
air showers, measured by AKENO\cite{halzen:94}. Finally, the
calculation labeled Zas (A) represents an extreme and crude upper
bound for the prompt fluxes, based on the assumption that a charm
is produced $10\%$ of the time in these high-energy interactions.
Actually, it is ruled out by the bounds set to the charm cross
section from the above mentioned analysis of AKENO data. 

One must be careful when analyzing differences in the curves of
Figure~\ref{fig:reported}. We call attention to the
fact that not only different charm models are being compared. The
reported prompt fluxes are calculated adopting different
atmospheric cascading routines. In order to better evaluate the
observed discrepancies, we proceed the analysis by taking into
account the separate effect of each single ingredient in the
shower process. Later on, we analyze the effect of multiple
ingredients, to evaluate how the choice of a different combination may
affect the final result.

\subsection{Single ingredient effects}
If we calculate the fluxes of prompt neutrinos ($\nu_{\mu}$ 
and $\nu_{e}$), we obtain essentially the same values as the
prompt muon flux (see for example 
Refs.~\cite{bugaev:98,tig:96,ggv:00}). 
The reason is that both the parent ($D$ or $\Lambda_{c}$)  and the
daughter ($K$ or $\Lambda_{o}$) particles are massive compared to
the leptons and the decay kinematics become blind to lepton family
number or flavor.

The prompt lepton flux is also essentially independent of $x$ (for
depths greater than a few interaction lengths),  due to the fact
that the main contribution to the high-energy lepton flux must
come from the first interactions of primary nucleons with air
nuclei, while they are still energetic enough. In addition, for a
fixed detection level, the flux is insensitive to the zenith angle
$\theta$, up to the charm critical energy.  This is also a
consequence of the fact that the bulk of energetic leptons are
produced high in the atmosphere and will reach the detector,
regardless the amount of atmosphere traversed (less depth in the
vertical direction, larger slant depths for showers close to the
horizon).  For energies above the critical energy, the charm
particle decay length becomes comparable to the interaction
length, and we feel the effects of its angular dependence, given
in Eq.~(\ref{eq:decay}). 

Since the prompt lepton flux is almost independent of lepton
flavor, detection depth and zenith angle, hereafter we will
perform our comparisons using the muon-neutrino vertical flux
($\theta = 0^{o}$) at sea level ($x\approx 1000$ g/cm$^{2}$).

\begin{figure} 
\begin{center}
 \includegraphics[height=11cm,angle=-90]{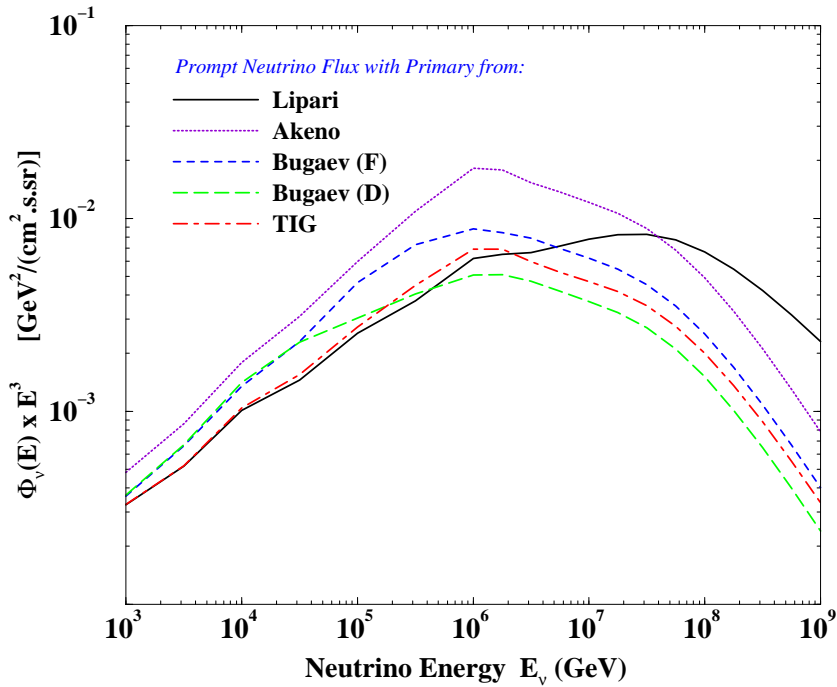}
\end{center}
\caption[]{Comparison of prompt neutrino fluxes  
for different primary spectrum models (see Table~\ref{table:primes}), 
assuming all the other ingredients fixed.}
\label{fig:result1}   
\end{figure}

First we investigate the effect of the primary spectrum at the top
of  the atmosphere (Section~\ref{section:primes}). 
To do so we calculate the prompt lepton flux, keeping all
ingredients fixed, but the primary flux. Just for comparison
purposes, we consider ingredients for the showering process used
by Bugaev. 
The effect of the primary spectrum, alone, to the final result is
considerable, as shown in Figure~\ref{fig:result1}. 
The spread on the resulting fluxes generally
increases with energy. 
For example, at $10^{9}$GeV, the difference between using Lipari's
single slope and TIG's primary is a factor 10 times, while they
started together at $10^{3}$ GeV. 
Also a big shift is present for the curve with AKENO primary
against all others with knee, above $10^{5}$ GeV.

The next two ingredients considered are the nucleonic interaction length and 
the nucleonic Z-moment (Section~\ref{section:sigmas}).  Apart from assuming a 
constant interaction length (as done by Lipari), for which the
overestimated value at high energies pulls the neutrino flux down,
the resulting fluxes are rather insensitive to the choice of
$\lambda_{N}$. Similar situation occurs with the fluxes calculated
changing only the nucleonic Z-moments, the difference being that
the overestimated flux comes from TIG ($Z_{NN}(\gamma) \approx
0.5$, at energies below $10^{5}$ GeV), pushing the neutrino flux
up, while other models ($Z_{NN}(\gamma) \approx 0.2 - 0.3$),
result basically in the same final prompt flux. 

The fluxes are also rather insensitive to the charm interaction
length up to $10^{7}$ GeV, as they should, since that is about the
value of $\varepsilon_{critic}$ for charmed particles (see
Table~\ref{table:pdg}). Above this range we discriminate the
models up to a factor of two times, whether we use $\lambda_i$
constant or with a $\log(E)$ dependence. 

\begin{figure} 
\begin{center}
 \includegraphics[height=11cm,angle=-90]{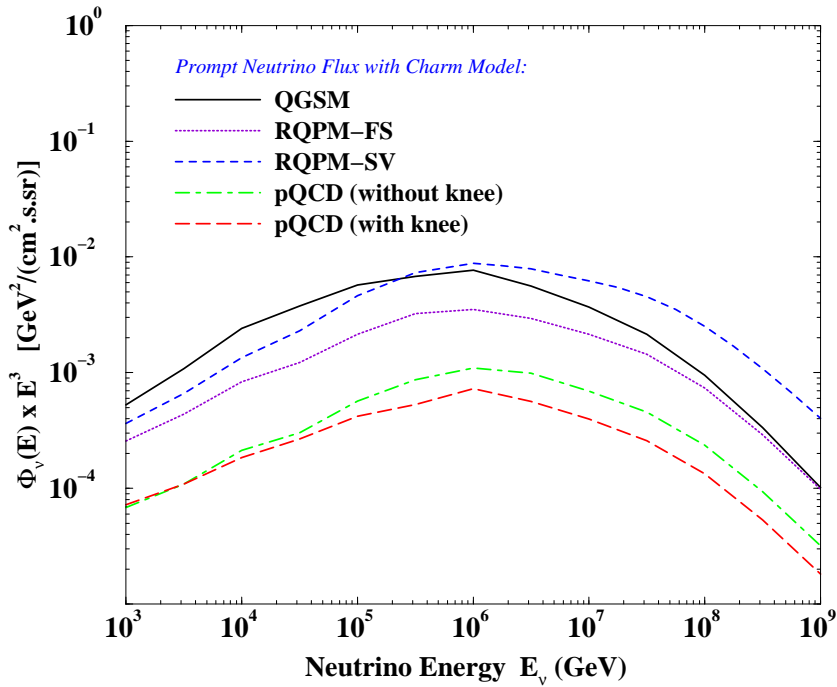}
\end{center}
\caption[]{Comparison of prompt neutrino fluxes  
for different charm Z-Moment models (see Section~\ref{section:zch}), 
assuming all the other ingredients fixed.}
\label{fig:result2}   
\end{figure}

When evaluating the effects of the choice of charm production
model (Section~\ref{section:zch}), the big uncertainties in
the inclusive cross sections of charm production are transmitted
to the calculated prompt lepton fluxes, as seen in
Figure~\ref{fig:result2}. 
The spread between the prompt flux calculated with RQPM-SV and
pQCD reaches a multiplicative factor of 20 at higher energies,
solely due to the choice of $Z_{Ni}(\gamma)$.

\subsection{Extreme ingredient combinations}
\begin{figure} 
\begin{center}
 \includegraphics[height=11cm,angle=-90]{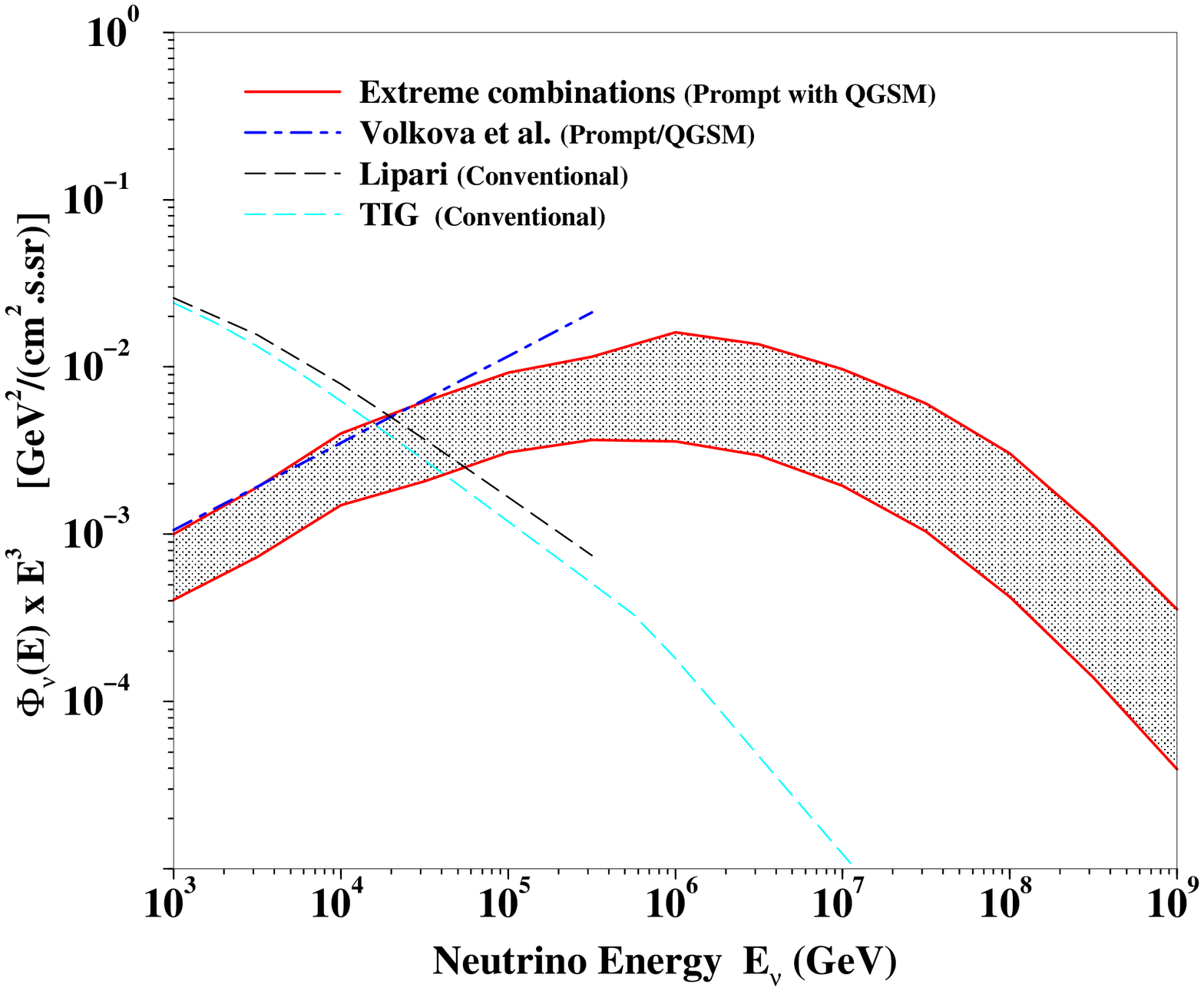}
\end{center}
\caption[]{Prompt neutrino flux calculated with QGSM. 
The band represents the range between extreme ingredient combinations. 
Prompt flux from Volkova {\it et al.} and 
conventional neutrino fluxes also shown for illustration.}
\label{fig:result3}   
\end{figure}

\begin{figure} 
\begin{center}
 \includegraphics[height=11cm,angle=-90]{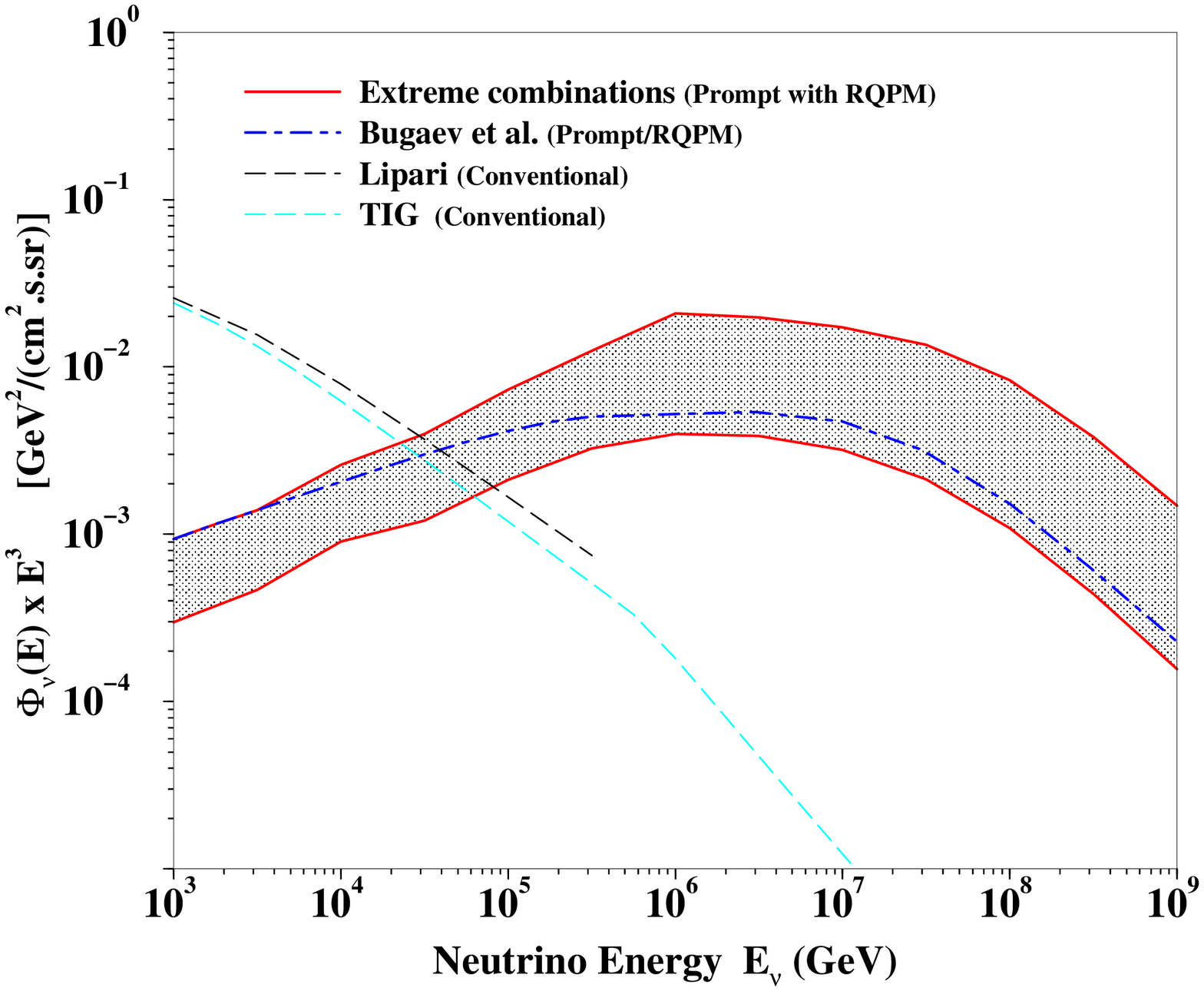}
\end{center}
\caption[]{Prompt neutrino flux calculated with RQPM. 
The band represents the range between extreme ingredient combinations. 
Prompt flux from Bugaev {\it et al.} and 
conventional neutrino fluxes also shown for illustration.}
\label{fig:result4}   
\end{figure}

\begin{figure} 
\begin{center}
 \includegraphics[height=11cm,angle=-90]{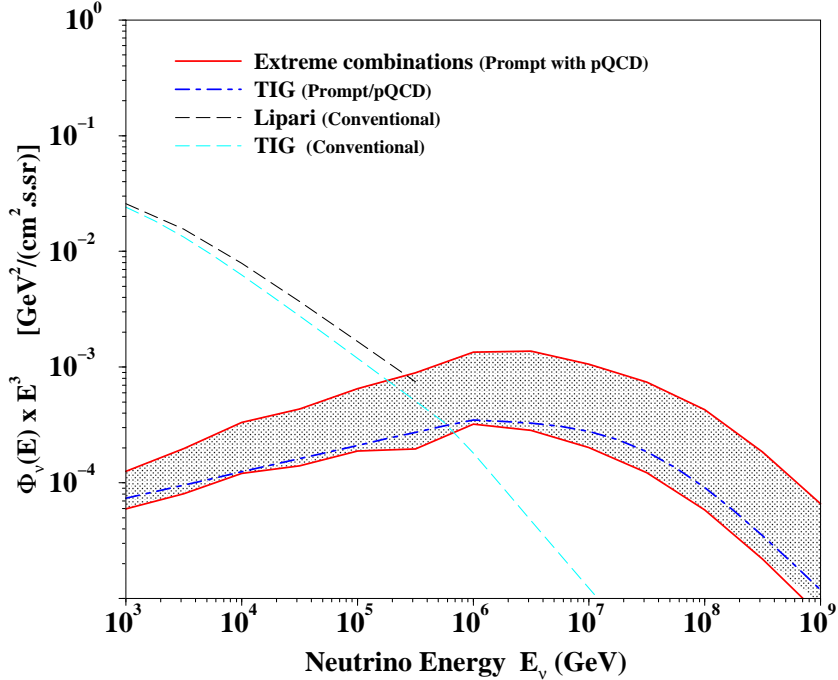}
\end{center}
\caption[]{Prompt neutrino flux calculated with pQCD. 
The band represents the range between extreme ingredient combinations. 
Prompt flux from TIG and 
conventional neutrino fluxes also shown for illustration.}
\label{fig:result5} 
\end{figure}

In the previous section the effect of each individual ingredient 
of the atmospheric showering process has been appreciated. Now, 
we investigate how the choice of multiple ingredients
combined in different ways affects the resulting flux. Let's
consider an example. The choice of the  production model QGSM
makes the flux dominates over other choices of charm production
only up to about $2 \times 10^5$~GeV, if all other parameters are
fixed (Figure~\ref{fig:result2}). The combination of the QGSM with
the AKENO primary flux, which  solely contributes to larger prompt
flux up to $10^{7}$~GeV  (Figure~\ref{fig:result1}),  makes this
new set dominate upon a broader range of energy. Choosing a
charm interaction length model that is responsible for increasing
the prompt flux at higher energies, we build a set which
dominates over yet a larger range. Such a combination of
showering parameters leads to a maximum flux configuration.
With a similar procedure we can build a minimum configuration.  
In this case it is interesting to consider the effect of
Lipari's primary up to the energy of the knee, above which
Bugaev's Model-D for the primary spectrum produces the minimum
prompt flux. The other showering parameters may also be chosen as
to stress the maximum or minimum possible behavior of a given flux
calculation.  Table~\ref{table:combinations} presents our
suggestion of multiple ingredient combinations selected to obtain
extreme outputs.  The prompt lepton flux curve calculated for each
charm production model (QGSM, RQPM or pQCD), can be shifted up or
down, depending on the chosen combination of ingredients, an
effect illustrated in Figures~\ref{fig:result3}
to~\ref{fig:result5}.  The band in each figure reflects the
freedom to change the resulting flux between the maximum (MAX) and
minimum (MIN) extreme combinations listed in
Table~\ref{table:combinations}. 

\begin{table}[ht]  
\caption[]{\label{table:combinations}
Extreme ingredient combinations of atmospheric showering
parameters, selected to obtain maximum or minimum behavior of
prompt flux.} 
\smallskip
\centering
\def\arraystretch{1.25}
\begin{tabular}{lcc} 
\hline\hline 
Ingredient  & MAXIMUM Set& MINIMUM Set\\ \hline 
$N_{0}, \gamma$ (for $E<E_{knee}$) & AKENO & Lipari \\
$N_{0}, \gamma$ (for $E>E_{knee}$) & AKENO & Bugaev-D \\
$\lambda_{N} (E)$ & power-law\cite{akeno:84} 
   & constant\cite{lipari:93} \\ 
$Z_{NN}(\gamma)$ & scaling violation\cite{tig:96}  
   & constant with knee\cite{chs:95} \\  
$\lambda_{i} (E)$ & constant\cite{volkova:85} 
   & $\log(E)$\cite{bugaev:98} \\  
\hline\hline    
\end{tabular}  
\end{table}  

%
%
%

\section{Conclusion}
The calculation of the prompt lepton flux produced in the atmosphere 
by the semileptonic decay of charmed particles is rather straightforward, 
but we cannot say the same for the analysis of the results. 
For instance, the lack of precise information on high-energy 
charm inclusive cross-section in hadron-nucleus collisions is
accompanied by a variety of options for the
particle showering process in the atmosphere. We described 
different ingredients of the calculation, comparing side by side
several parametrizations for each one of them, and evaluated their
relative importance to the final result. 
The major effects are due to the choice  of the primary
spectrum at the top of the atmosphere and, of course,  to the
choice of the charm particle production model.  Only first
nucleonic interactions play essential role in determining the
prompt lepton flux down at sea level,  therefore variations in
the nucleonic attenuation lengths are not that relevant, while
the charm interaction length have some influence, above the charm
particle critical energy. 
We observed how different combinations of ingredients can shift
the resulting flux curves up or down. Therefore, the comparison
between different calculations must be carried out with great
care.  The bands displayed in 
Figures~\ref{fig:result3}-\ref{fig:result5} correspond to 
the allowed regions for the prompt lepton flux calculated
respectively with the charm production models QGSM, RQPM and
pQCD, if the showering process parameters are mixed differently. 

The prompt lepton crossover energy,  that is the energy
above which the charm particle decay products dominate  over the
conventional pion and kaon decay induced fluxes, is yet an
uncertain quantity. 
According to Figures~\ref{fig:result3}-\ref{fig:result5}, 
it may be anywhere between $10^{4}$ and $10^{6}$ GeV. 
If the cross-over is ``low enough'' (about $10^{4}$ GeV), then 
neutrino telescopes now operational can therefore take advantage
of the isotropy of the prompt lepton flux, to search for an 
zenith angle independent component in their data. This can also be
pursued by the analysis of the more copiously detected down-going
muons. 

Exploiting the case of tau-neutrinos, which may produce a clear signature in
high-energy neutrino detectors\cite{learned:95}, will be addressed in a
future analysis. There is also the need for a more comprehensive description
of the available data on charm production cross section, and its extrapolation 
to higher energies. 

The prompt lepton flux is on the order-of-the-day of operating high-energy
neutrino telescopes, because of the background it represents.
Proposed experiments, like IceCube~\cite{ice3:00}, 
may turn the arguments the other way around, 
for their measurements with enhanced sensitivity  
may provide outstanding information on heavy quark interactions, 
just by discriminating atmospheric from cosmic neutrinos, 
at energies above tens of TeV.

\section*{Acknowledgements}
The author would like to thank Freddy Binon, Jean-Marie Fr\`ere,  
Francis Halzen and C. Salles for many discussions and suggestions
to the manuscript. This work was partially supported by the
I.I.S.N. (Belgium) and by the Communaut\'e Fran\c{c}aise de
Belgique - Direction de la Recherche Scientifique, programme ARC.

\end{document}